\documentclass[review]{elsarticle}

\usepackage{lineno,hyperref}
\modulolinenumbers[1]

\usepackage{color}

\journal{Journal of \LaTeX\ Templates}

\begin{document}

\begin{frontmatter}

\title{Diagnostics for ultrashort X-ray pulses using silicon trackers}

\author[PRI,DEP,PPL]{Jiaxing Wen}

\author[PPL]{Minghai Yu}

\author[PPL]{Yuchi Wu}

\author[PRI,DEP]{Ming Zeng\corref{correspondingauthor}}
\cortext[correspondingauthor]{Corresponding author}
\ead{zengming@mail.tsinghua.edu.cn}

\author[PPL]{Bo Zhang}

\author[DA,DEP]{Jirong Cang}

\author[DEP]{Yuge Zhang}

\author[DEP]{Ge Ma}

\author[PPL]{Yue Yang}

\author[DEP,PPL]{Wenbo Mo}

\author[PPL]{Zongqing Zhao}

\address[PRI]{Key Laboratory of Particle and Radiation Imaging (Tsinghua University), Ministry of Education, Beijing 100084, China}
\address[DEP]{Department of Engineering Physics, Tsinghua University, Beijing 100084, China}
\address[PPL]{Science and Technology on Plasma Physics Laboratory, Laser Fusion Research Center, CAEP, Mianyang 621900, Sichuan, China}
\address[DA]{Department of Astronomy, Tsinghua University, Beijing 100084, China}

\begin{abstract}
    The spectrum of laser-plasma generated X-rays is very important, it characterizes electron dynamics in plasma and is basic for applications. However, the accuracies and efficiencies of existing methods to diagnose the spectrum of laser-plasma based X-ray pulse are not very high, especially in the range of several hundred keV. In this study, a new method based on electron tracks detection to measure the spectrum of laser-plasma produced X-ray pulses is proposed and demonstrated. Laser-plasma generated X-rays are scattered in a multi-pixel silicon tracker. Energies and scattering directions of Compton electrons can be extracted from the response of the detector, and then the spectrum of X-rays can be reconstructed. Simulations indicate that the energy resolution of this method is approximately $\rm 20\%$ for X-rays from $\rm 200$ to $\rm 550$ keV for a silicon-on-insulator pixel detector with 12 $\rm \mu$m pixel pitch and 500 $\rm \mu$m depletion region thickness. The results of a proof-of-principle experiment based on a Timepix3 detector are also shown.
\end{abstract}


\end{frontmatter}


\section{Introduction}
\label{sec:intro}
X-ray sources have many applications in fields such as radiography\cite{ISI:000468859400011}, nondestructive testing (NDT)\cite{jones_evaluating_2016}\cite{yang_design_2019} and plasma hydrodynamic diagnostics\cite{tian_radiography_2019}\cite{weimin_high-resolution_2020}. Laser-plasma-based X-ray sources have attracted considerable attention in recent years owing to their compact size, short duration and broadband spectrum\cite{edwards_characterization_2002}\cite{yu_ultrahigh_2016}\cite{lemos_bremsstrahlung_2018}\cite{wang_brilliant_2020}. Characterisation of laser-plasma based X-ray sources, e. g., their spectrum, is then very important, as it is useful for applications and monitoring electron dynamics in plasma\cite{kneip_observation_2008}.

There are many existing detectors that can measure X-ray spectrum, such as gas detectors, scintillation detectors, and semiconductor detectors\cite{Knoll2010Radiation}. Most of these conventional X-ray detectors measure the energies of X-ray photons one by one to give the spectrum, hence are applicable only in low fluence rate cases. However, laser-plasma-based X-ray sources have extremely high fluence rate, usually emit $\rm > 10^8$ photons within a duration of $\rm 10$s of femtoseconds, which is much shorter than the typical dead time of these conventional detectors. Hence diagnosing laser-plasma generated X-ray pulses is challenging. Many techniques, e.g., diffraction-based\cite{chi_diffraction_2017}\cite{young_high-resolution_1998}\cite{yu_hard_2016}, filter-based\cite{rusby_novel_2018}\cite{behm_spectrometer_2018}\cite{jeon_broadband_2015}\cite{fehl_spectral_2000}\cite{minghai_development_2017}, Compton-scattering-based\cite{singh_compact_2018}\cite{quiter_developing_2018}\cite{espy_wide-acceptance_2016}\cite{schumaker_measurements_2014}, and single-photon-counting-based\cite{yan_calibration_2013}\cite{stoeckl_operation_2004}, have been proposed and applied in measurements of the spectrum of ultrashort X-ray pulses. These methods based on different mechanisms are suitable for different energy bands of X-rays.

Diffraction-based methods usually use crystals to diffract X-rays, and the spectrum can be reconstructed from the angular distribution of diffracted X-rays. The diffraction-based methods have high spectral resolution for comparatively low energy X-rays, usually below $\rm 100$ keV, but suffer from the low diffraction efficiency. Compton-scattering-based methods convert the X-ray spectrum into electron spectrum which can be measured easily. The detection efficiencies of Compton-scattering-based methods are comparatively low, this is due to the generation and collimation processes of Compton electrons. Filter-based methods reconstruct the X-ray spectrum in a broadband from the deposited energy in a stack of filters. However, such reconstructions suffer from ill-posed linear system therefore have comparatively high errors. Single-photon-counting-based methods can provide high-resolution X-ray spectrum using pixel detectors, e. g., scientific charge-coupled devices (CCD), operated in single-photon counting mode. However, high energy X-ray photons, typically above $\rm 100$ keV, only deposit part of their energies in the sensor, which results in inaccurate measurement of photon energies. Hence there is an apparent need for accurate, high-efficient spectrum measurement of ultra short laser-plasma based X-ray pulses of several hundred keV.

Recently, some new methods based on electron track detection have been proposed \cite{zhang_ccd-based_2014}\cite{zhang_diagnostics_2016}\cite{quiter_developing_2018}\cite{claps_timepix3_2019} to diagnose the spectrum of ultrashort gamma pulses of $\rm 1-100$ MeV with better efficiency and fast post-shot processing. \textcolor{black}{Zhang and Quiter use a thin scatter foil to convert the incident photon beam of several MeV to energetic Compton electrons, and the Compton electrons escape from the foil and are captured by CCDs. The energies and directions of the Compton electrons are estimated from the electron tracks in the CCDs, and then the energy of photons can be calculated from Compton formalism\cite{zhang_ccd-based_2014}\cite{zhang_diagnostics_2016}\cite{quiter_developing_2018}. Claps use a Timepix3 detector to detect the gamma photons directly, and the characteristic energy of photon beams is estimated by machine learning methods from the physical and morphological parameters of Compton electron tracks\cite{claps_timepix3_2019}. These studies are inapplicable to diagnose the X-ray pulses of several hundred keV but show the potential to diagnose high fluence rate X-ray pulses with electron track detection.}

In this report, an X-ray spectrometer based on detection of Compton electron tracks that can be applied to laser-plasma produced X-ray pulses is presented. Laser-plasma generated X-rays are scattered in the sensor of a multi-pixel silicon detector and its response is recorded as a gray level image. After that, the image is processed by a clustering algorithm and an electron track algorithm to extract Compton electron tracks and calculate the scattering directions of Compton electrons. The spectrum of incident X-rays can be reconstructed from these tracks. Simulations show that X-rays from $\rm 200$ to $\rm 550$ keV can be measured with $\rm \sim 20\%$ energy resolution with a silicon-on-insulator (SOI) detectors\cite{miyoshi_monolithic_2013} with $\rm 500$ $\rm \mu$m thick sensor and $\rm 12$ $\rm \mu$m pixel pitch. A proof-of-principle experiment conducted by irradiating a Timepix3 detector with a radioactive source is also discussed.

\section{Methodology}
\label{sec:CD}
\subsection{Compton scattering}
\begin{figure}[!htb]
\centering
	\includegraphics[width=0.8\textwidth]{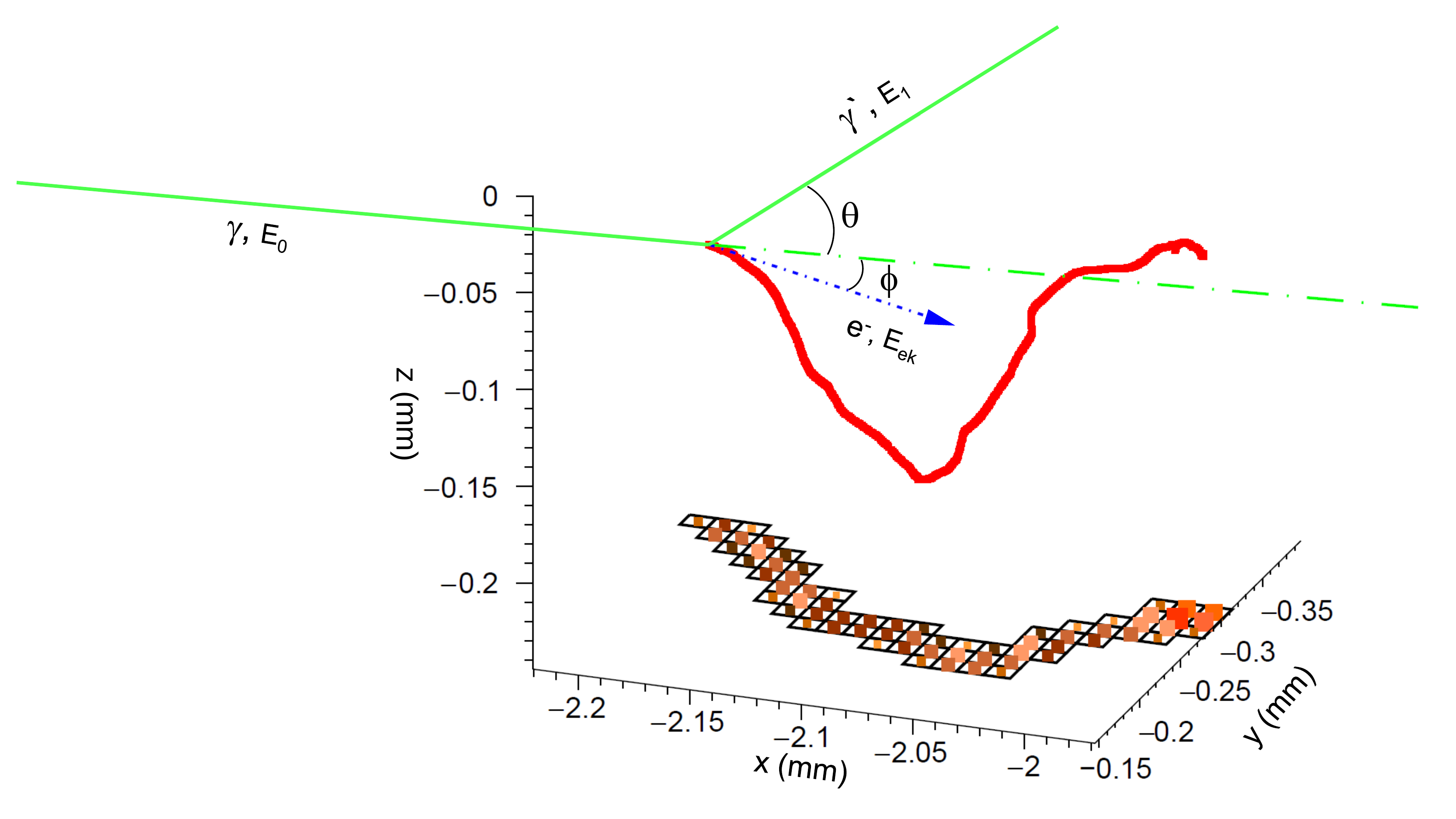}
	\caption{Compton scattering and detection schematic. Photon (green solid) and electron (red solid thick) trajectories, incident photon (green dot-dashed) and scattered electron (blue dot-dashed) directions obtained in a Monte Carlo simulation are shown in the upper half. The borders of triggered pixels are shown below the electron trajectory, and the size and colour of squares at the centre of the pixels represent the signal of the pixels.}
	\label{Fig:ComptonScatter}
\end{figure}

For X-rays with energies of several hundreds keV, the main interaction mechanism between photons and matter is Compton scattering. When a photon transfers some part of its energy to an electron, the energetic electron propagates through the detector and triggers many pixels along the electron trajectory as shown in Fig.\ref{Fig:ComptonScatter}.

In Fig.\ref{Fig:ComptonScatter}, an incident photon $\rm \gamma$ with energy $\rm E_0$ is scattered incoherently in the sensor of a pixel detector, producing a Compton photon $\rm \gamma'$ with energy $\rm E_1$ and a Compton electron with kinetic energy $\rm E_{ek}$. The scattering angles of the Compton photon and Compton electron are $\rm \theta$ and $\rm \phi$, respectively. According to the Compton formalism,
\begin{equation}
\rm E_0 = \frac{m_0c^2}{\sqrt{1+\frac{2m_0c^2}{E_{ek}}}cos\phi-1},
\label{eq:photon_enengy}
\end{equation}
$\rm E_0$ can be calculated from $\rm \phi$ and $\rm E_{ek}$, where $\rm m_0$ is the mass of the stationary electron, and c is the speed of light. $\rm E_{ek}$ can be deposited in the detector and read out directly. The interaction point and scattering direction of the Compton electron can be reconstructed from the response of the pixel detector using the electron track algorithm\cite{li_electron_2017}. The direction of incident X-ray is approximately determined by the point of impact (POI) of the laser and the reconstructed interaction points in the detector.

When an X-ray pulse interacts with a pixel detector, many Compton electrons are produced. Their tracks can be extracted using a clustering algorithm and their energies and scattering directions can be calculated from their tracks. Then the spectrum of laser-plasma-induced X-rays can be reconstructed.

\subsection{Compton electron track characteristics and detector}
\label{sec:cetc}

To determine what kind of detector would be suitable, properties of Compton electrons were investigated. The angular distribution of Compton scatter from free and stationary electrons is given by the Klein–Nishina differential cross-section, from which the cross section for the number of electrons scattered between two cones of half-angles $\rm \phi$ and $\rm \phi + \delta\phi$ can be derived\cite{davisson_gamma-ray_1952}:

\begin{eqnarray}
\rm\frac{d\sigma_e}{d\phi}&&\rm{ = \frac{r_0^2}{2}\{\frac{E_1}{E_0}+(\frac{E_1}{E_0})^3+(\frac{E_1}{E_0})^2[(\frac{m_ec^2}{E_1}-\frac{m_ec^2}{E_0}-1)^2-1]\}\nonumber} \\
&&\rm{\times\frac{4(\frac{E_0}{m_ec^2}+1)^2\cos(\phi)\times2\pi\sin(\phi)}{[(\frac{E_0}{m_ec^2}+1)^2-\frac{E_0}{m_ec^2}(\frac{E_0}{m_ec^2}+2)\cos^2(\phi)]^2}}.
\label{eq:crosssection}
\end{eqnarray}

\begin{figure}[!htb]
\centering
	\includegraphics[width=0.8\textwidth]{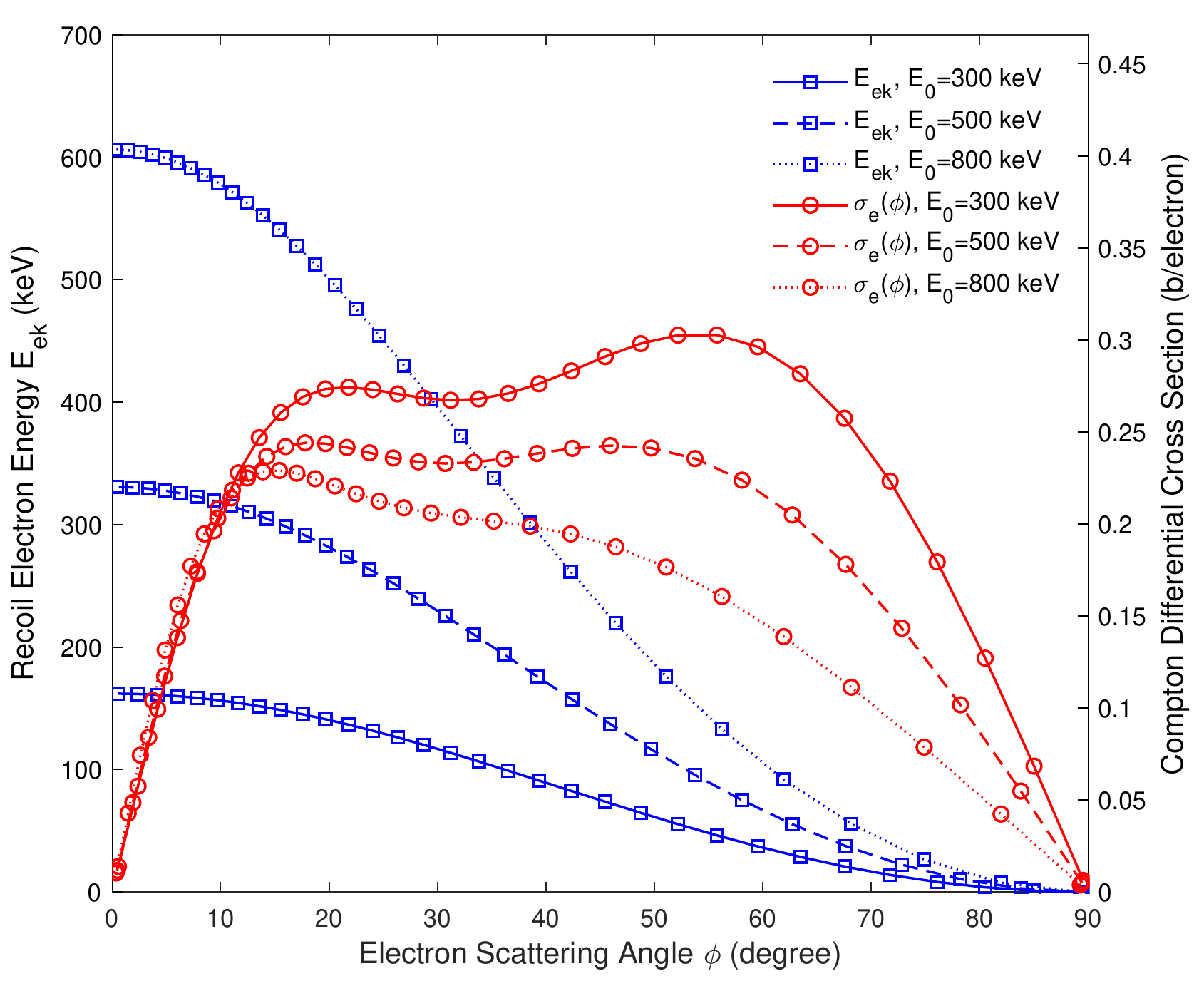}
	\caption{Energies of Compton electrons and differential cross section per unit angle for the number of electrons scattered in the direction $\rm \phi$.}
\label{Fig:ElectronEnergyAndCrossSection}
\end{figure}

Fig.\ref{Fig:ElectronEnergyAndCrossSection} shows the distribution of energy and cross section $\rm \frac{d\sigma_e}{d\phi}$. For incident photons of hundreds of keV, Compton electrons have wide angular distribution and their energies are of few hundreds keV scale.

Then let us analyze restrictions on the sensor. First, according to Eq.(\ref{eq:photon_enengy}), calculation of the incident photon energy need $\rm E_{ek}$. To precisely measure $\rm E_{ek}$, the sensor should make the Compton electrons deposit their energies as much as possible. Hence this diagnostic method prefers larger sensor and denser sensor material. Second, to give the scattering direction of Compton electron, which is also needed for calculate the $\rm \gamma$ energy $\rm E_0$ according to Eq.(\ref{eq:photon_enengy}), we need their trajectories as precise as possible. Therefore smaller pixels are preferred. Third, more than hundreds of electron tracks that do not overlap with each other are needed in a single shot to provide enough samples for the X-ray spectrum reconstruction. This statistical requirement prefer smaller pixels and larger senors.

Gaseous trackers show good angular resolution and have been widely used for electron track detection\cite{feng_polarlight_2019}\cite{takada_development_2005}. However, for electrons with energies above $\rm 100$ keV, their tracks in gaseous trackers would be too long. They cannot deposit all their energies and the tracks would easily overlap. Electron tracks are much shorter in solid scintillation trackers, but their angular resolutions are usually poor due to the dispersion of scintillation light and large pixel pitches. Semiconductor trackers, such as scientific CCD\cite{plimley_reconstruction_2011}, SOI pixel detectors\cite{miyoshi_monolithic_2013} and hybrid detectors\cite{poikela_timepix3_2014}, are much more suitable. They usually consist of a large number of small pixels, the pixel pitches of available silicon-based semiconductor tracker can be as small as 10 $\rm \mu$m and their depletion regions can be as thick as 500 $\rm \mu$m. The stopping distance of $\rm 100$–$500$ keV electrons in silicon ranges from $\rm 78$ to $\rm 943$ $\rm \mu$m, therefore, most Compton electrons from $\rm 100$ to $\rm 500$ keV can deposit all their energies and generate recognisable tracks. As shown in previous study\cite{plimley_angular_2016}, the pixel pitch of a silicon-based semiconductor tracker should be less than 20 $\rm \mu$m for a reasonable angular resolution of hundreds of keV electrons, and there are only a few types of semiconductor trackers with two-dimensional (2D) spatial resolution that have such a small pixel pitch and thick depletion region. The scientific CCD\cite{plimley_reconstruction_2011} and the SOI pixel detector INTPIX5\cite{miyoshi_monolithic_2013} are examples.

\begin{figure}[!htb]
\centering
	\includegraphics[width=0.8\textwidth]{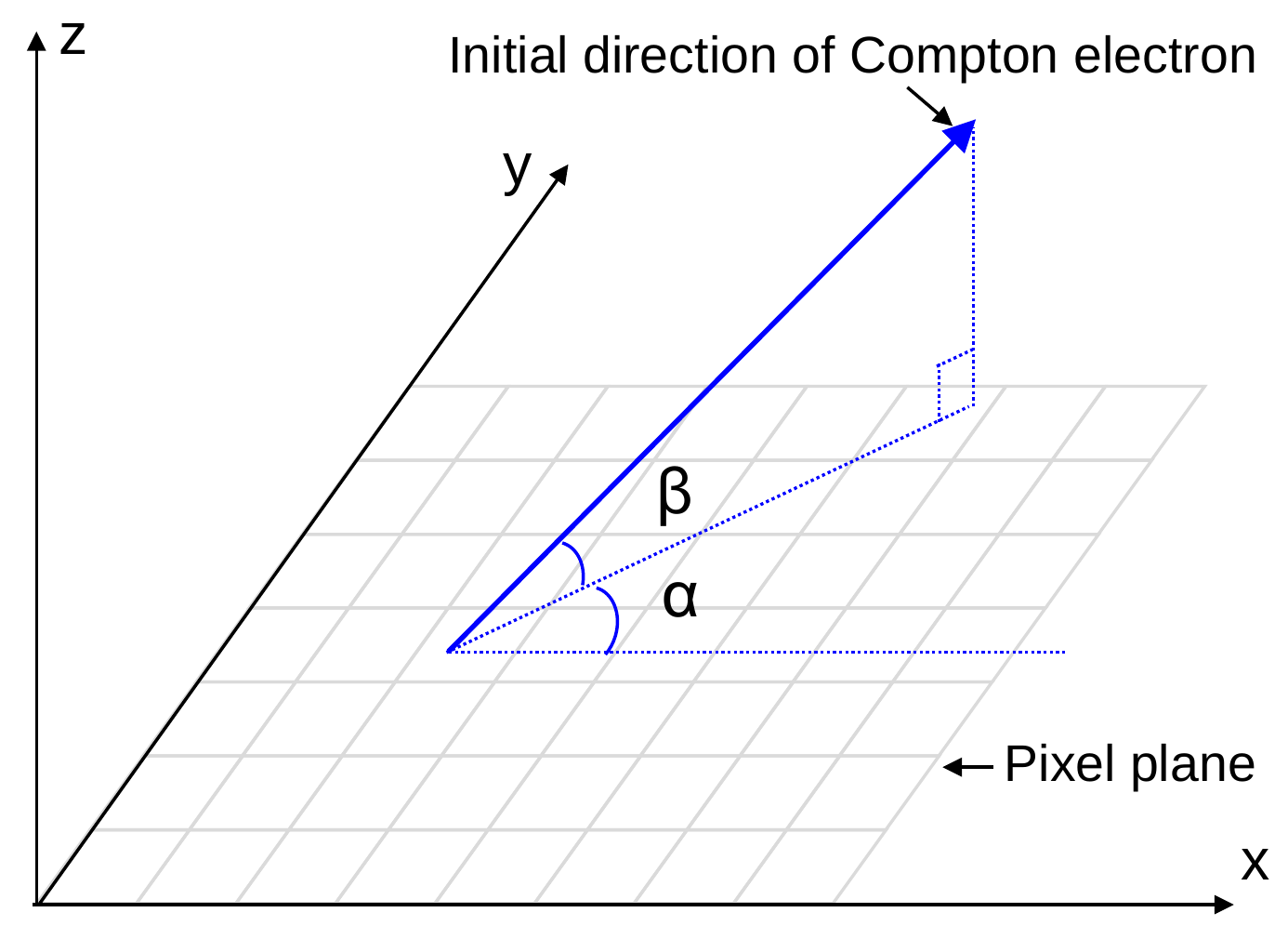}
	\caption{\textcolor{black}{Two angles that fix the initial direction of a Compton electron. $\rm \alpha$ fixes the projection of electron direction in the pixel plane and $\beta$ is the angle out of the pixel plane.}}
\label{Fig:AlphaAndBeta}
\end{figure}

For a $\rm 2$D semiconductor tracker, the initial direction of a Compton electron can be fixed by two angles, $\rm \alpha$ and $\rm \beta$ \cite{plimley_reconstruction_2011}\cite{plimley_experimental_2013}. \textcolor{black}{As shown in Fig.\ref{Fig:AlphaAndBeta},} $\rm \alpha$ fixes the projection of electron direction in the pixel plane, which can be measured with a CCD image; and $\beta$ is the angle out of the pixel plane, which can not be measured precisely and the sign of $\rm \beta$ is ambiguous \cite{plimley_reconstruction_2011}\cite{plimley_angular_2016}. When the incident direction of X-ray is perpendicular to the pixel plane, the scattering angle of Compton electron $\rm \phi$ is determined by $\rm \beta$. The imprecise measurement and ambiguous sign of $\rm \beta$ would make the reconstruction of X-ray spectrum difficult. Therefore, a $\rm 2$D pixel detector must be deployed as its pixel plane is parallel to the incident X-rays, i. e., the ``side-on'' configuration. The ``side-on'' configuration also allows the electron trajectories fully developed in the sensor body \cite{claps_timepix3_2019}.

\begin{figure}[!htb]
\centering
	\includegraphics[width=0.8\textwidth]{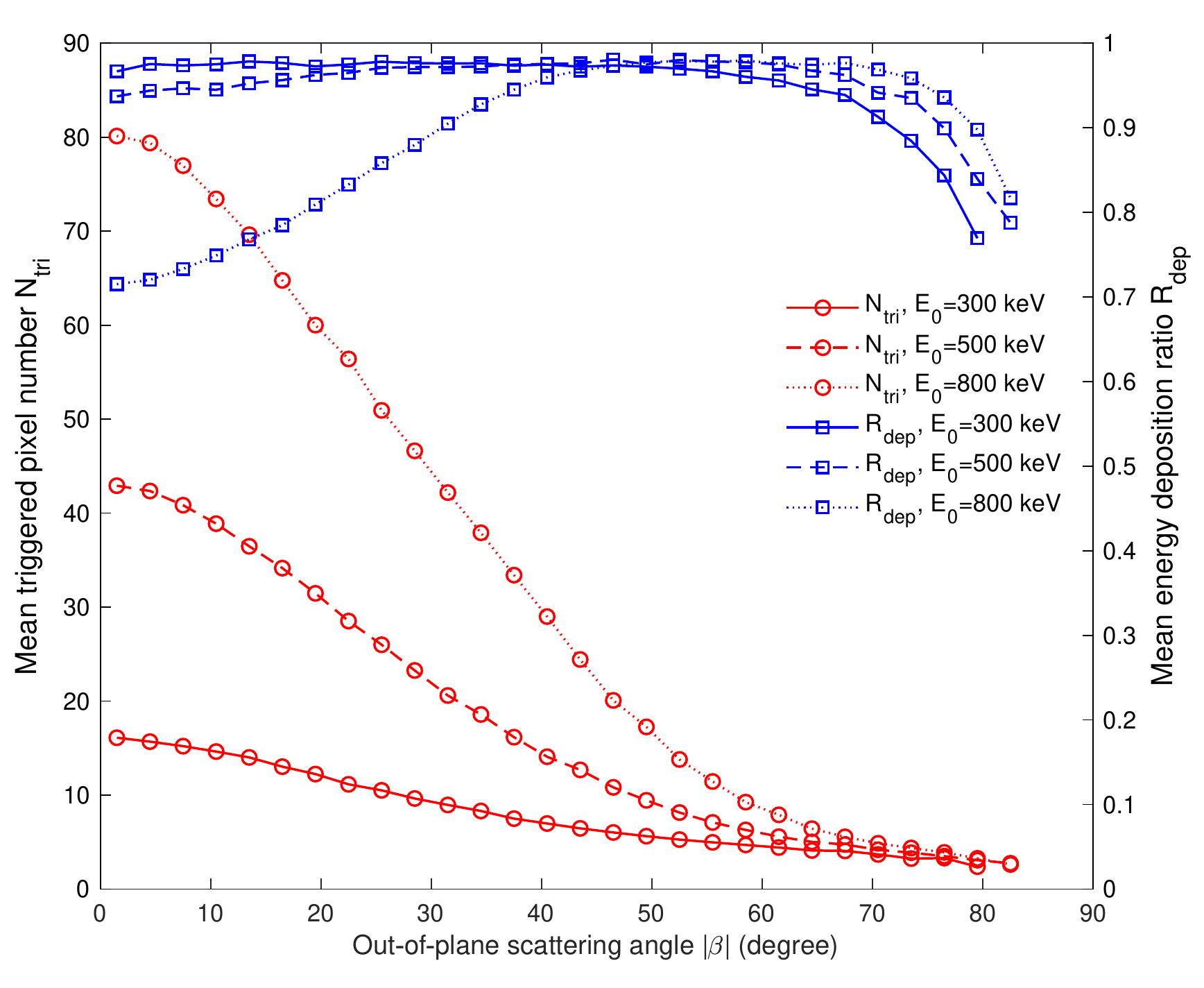}
	\caption{Dependence of mean energy deposition ratio and mean number of triggered pixels on $| \beta |$.}
\label{Fig:EnergyDepositionRatioWithBeta}
\end{figure}

The characteristics of electron tracks in a $\rm 2$D silicon tracker with $\rm 12$ $\rm \mu$m pixel pitch and 500 $\rm \mu$m depletion region thickness were studied using Monte Carlo simulations. The tracker was illuminated in the ``side-on'' configuration by a pencil beam with energies of $\rm 300$, $\rm 500$ and $\rm 800$ keV. The mean energy deposition ratio and mean number of triggered pixels with different $\rm | \beta |$ values are shown in Fig.\ref{Fig:EnergyDepositionRatioWithBeta}. Most Compton electrons with small $\rm |\beta |$ values can deposit most of their kinetic energies in the sensor and trigger a large number of pixels. Compton electrons with large $\rm | \beta |$ have a large scattering angle $\rm  \phi$ and a low energy and only trigger a few pixels. When $\rm | \beta |$ is close to 90$\rm ^\circ$, Compton electrons can easily escape from the sensor, and the energy deposition ratio become very low. Forward-directed, energetic electrons produced by back-scattering X-ray photons also tend to escape from the sensor, which result in a decline in the energy deposition ratio in small $\rm | \beta |$ regions.

For a $\rm 2$D pixel detector, obtained image can give the $\rm \alpha$ angle, while hard to reflect the value of $\rm \beta$ angle. This seems an obstacle for measuring Compton scattering angle $\rm \phi$, but fortunately, Compton electrons with large $\rm | \beta |$ angles have lower energies and trigger fewer pixels. Hence the number of triggered pixels decreases monotonically with $\rm | \beta |$ as shown in Fig.\ref{Fig:EnergyDepositionRatioWithBeta}. Therefore, simply discarding tracks with pixels numbers below a threshold can exclude large $\rm | \beta |$ events, \textcolor{black}{e.g., discarding the tracks with pixel numbers less than six.} In remaining events, $\rm \beta=0$ is a reasonable approximation for the Compton scattering directions.

\subsection{Electron track algorithm}
\begin{figure}[!htb]
\centering
	\includegraphics[width=0.95\textwidth]{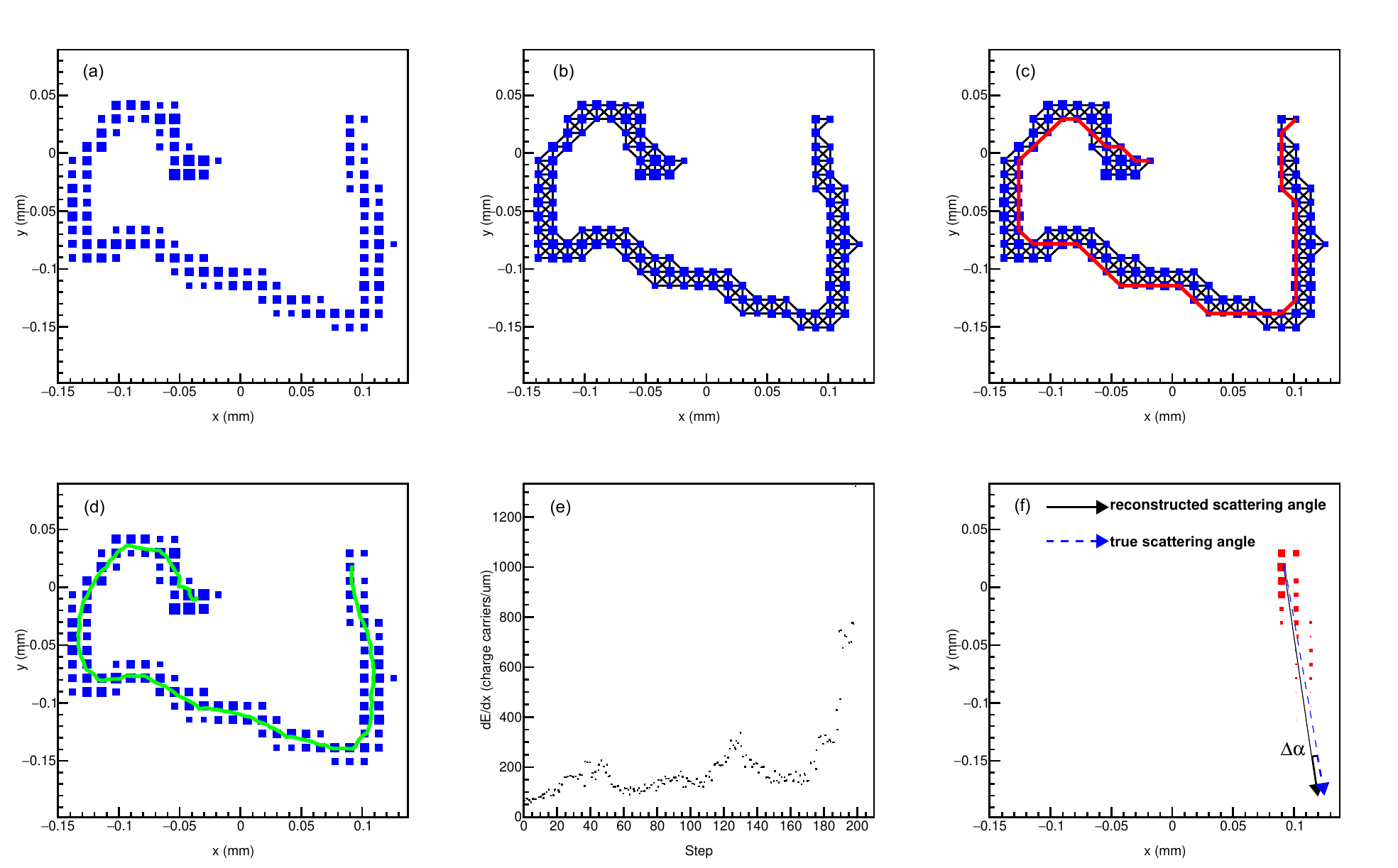}
	\caption{Illustration of the electron track reconstruction algorithm. (a) Simulated detector response. (b) Neighbouring points connected with solid lines. (c) The primary path (red line), defined as the longest one of all the shortest paths between every two points in the graph. (d) The reconstructed path (green line), derived from the primary path after the spatial energy filter. (e) The reconstructed path is divided into 200 steps, and the energies deposited in each step ,dE/dx, are estimated by calculating the sum of the pixel charge around each step. The Bragg peak is clear, and the start point (interaction point) and end point of the reconstructed path can be determined. (f) The 2D scattering direction of Compton electron derived as the direction of the maximum second moment of the distance-weighted charge map (red squares).}
\label{Fig:CSETA2D}
\end{figure}

Electron track algorithm can reconstruct the electron path, interaction point, and electron initial direction. The electron track algorithm based on the shortest path problem in graph theory, which had been successfully applied in the gaseous pixel detector for photoelectric X-ray polarimetry\cite{li_electron_2017} and pixelated CdZnTe detector for discrimination between single electron events and 0$\rm \nu\beta\beta$ events\cite{zeng_3-d_2017}, was adopted in this study. Steps of this algorithm are shown in Fig.\ref{Fig:CSETA2D}. The definition of the error of the reconstructed initial direction of electron $\rm \Delta\alpha$ is shown in Fig.\ref{Fig:CSETA2D}(f).

In the electron track algorithm, scattering direction of Compton electron is reconstructed as the direction of the maximum second moment of the distance-weighted charge map as shown in Fig.\ref{Fig:CSETA2D} (f). The weight factor we adopted is $\rm W(d_{ip})=exp(-d_{ip}/w)$ \cite{li_electron_2017}, where $\rm w$ is a constant, and the distance $\rm d_{ip}$ is the shortest distance from the pixel centre to the reconstructed path plus the distance along the path from start point to the closest point on the path. It can avoid the situation that the weight $\rm W(d_{ip})$ of pixels at the end of the electron track be larger due to electron track circling.

To evaluate the performance of this algorithm and derive the optimal $\rm w$ for silicon trackers with $\rm 12$ $\rm \mu$m pixel pitch, Monte Carlo simulations were conducted. In the simulations, electrons with energies from $\rm 50$ to $\rm 600$ keV were directly injected into the centre of the detector. The initial direction was defined by $\rm \alpha$ and $\rm \beta$, $\rm \beta$ was fixed to $\rm 0$, i. e., the incident direction is in the pixel plane, and $\rm \alpha$ follows a uniform distribution from $\rm 1$ to $\rm \pi$ rad.

\begin{figure}[!htb]
\centering
	\includegraphics[width=0.8\textwidth]{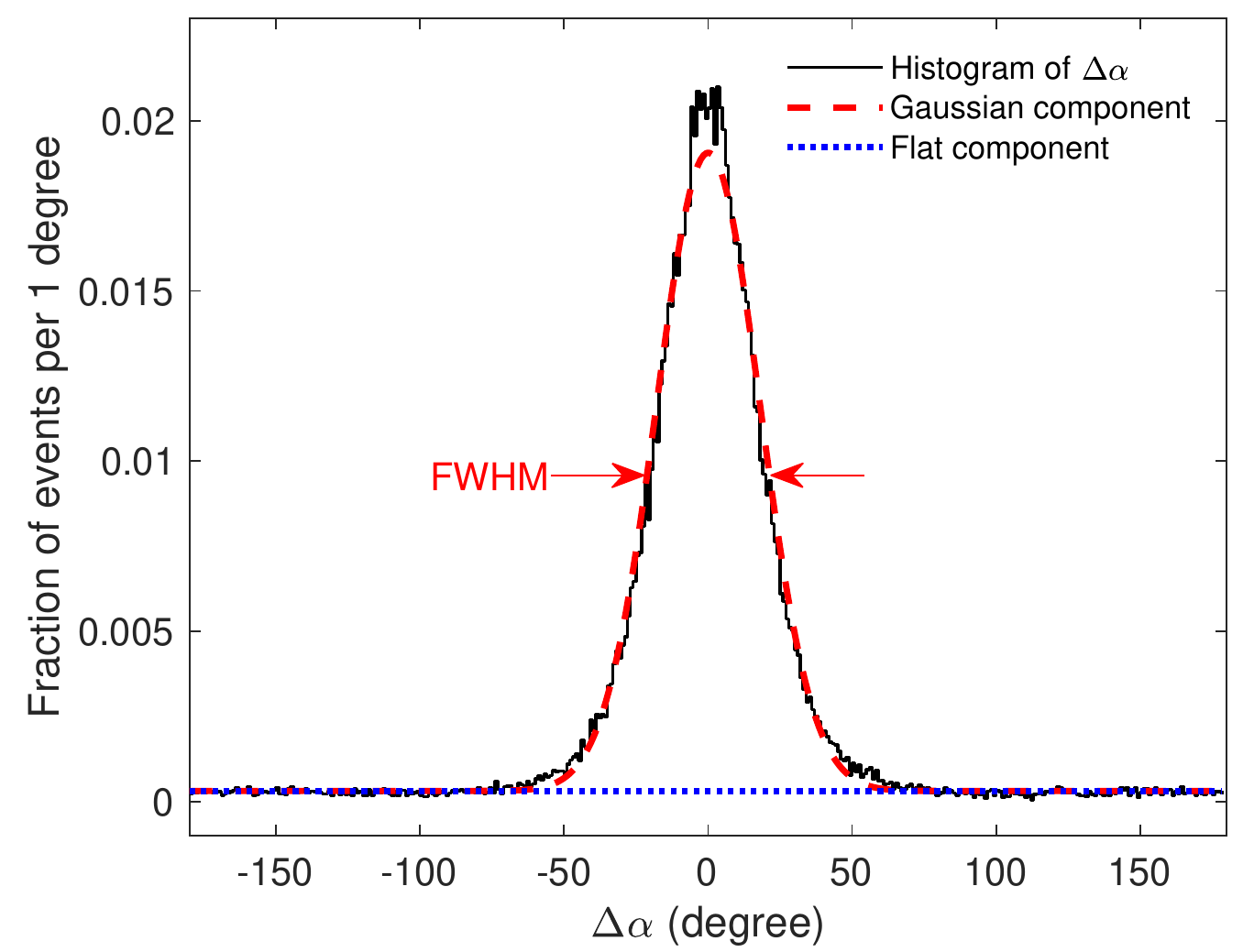}
	\caption{Distribution of $\rm \alpha$ measurement error for electrons with energy of 250 keV and $\rm w$ = 0.008 mm}
\label{Fig:Distribution_Alpha_Error}
\end{figure}

\begin{figure}[!htb]
\centering
	\includegraphics[width=0.8\textwidth]{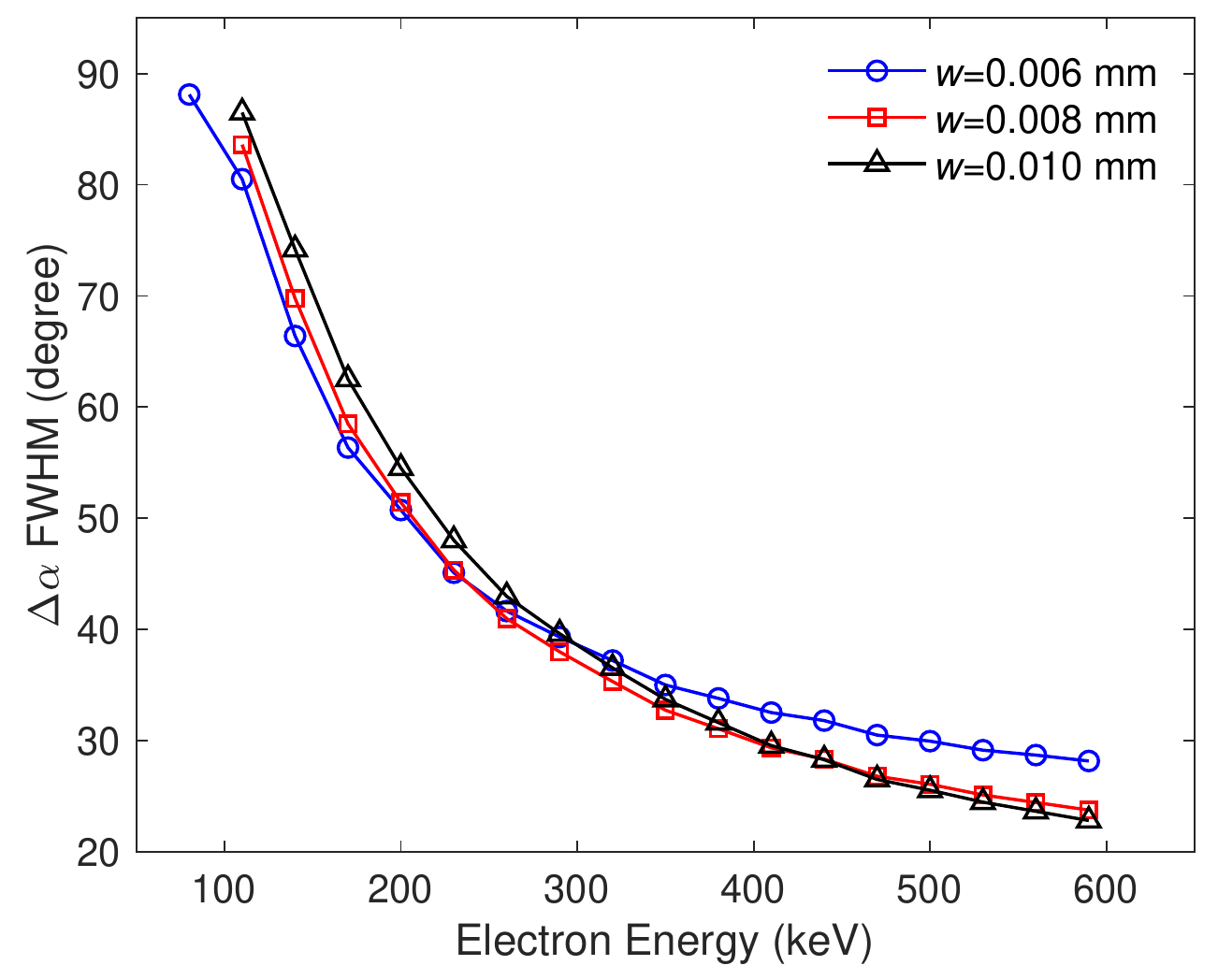}
	\caption{Dependence of $\Delta\alpha$ distribution width on electron energy.}
\label{Fig:FWHMDependence}
\end{figure}

The distribution of $\rm \Delta\alpha$ for $\rm 250$ keV electrons is shown in Fig.\ref{Fig:Distribution_Alpha_Error} as an example. The distribution of $\rm \Delta\alpha$ can be characterised as a Gaussian peak on a plateau \cite{plimley_angular_2016}. The dependence of the Full width at half maximum (FWHM) on electron energy and $\rm w$ is shown in Fig. \ref{Fig:FWHMDependence}. When electron energy is below $\rm 100$ keV, the $\rm \Delta\alpha$ distribution is very wide and the calculation of $\rm \alpha$ fails. As expected, the optimal $\rm w$ increases with electron energy, and the FWHM of $\rm \Delta\alpha$ distribution is within $\rm 30^\circ$ for electrons above $\rm 400$ keV with optimal $\rm w$, which is consistent with Ref. \cite{plimley_angular_2016}. Hence a dynamic $\rm w$ depending on the electron energy was adopted in this work.

\subsection{Diagnostic design}

\begin{figure}[!htb]
\centering
	\includegraphics[width=1\textwidth]{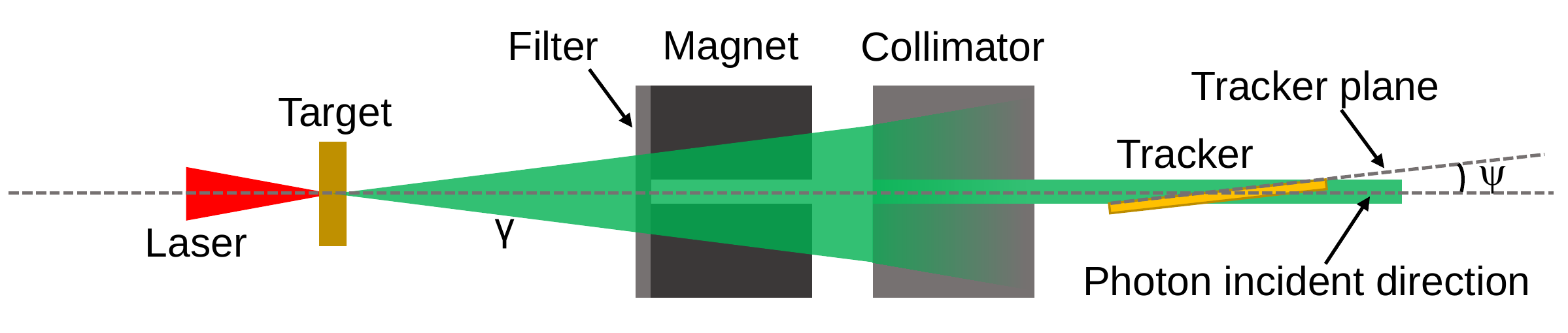}
	\caption{Conceptual design of the experiment.}
\label{Fig:ConceptualDesign}
\end{figure}

The conceptual design of the spectrum measurement setup is shown in Fig.\ref{Fig:ConceptualDesign}. Laser-plasma-induced X-rays are filtered, collimated, and incident on a silicon-based tracker. The angle between the X-ray incident direction and pixel plane $\rm \psi$ is $\rm 0$, i. e., the ``side-on'' configuration. A magnet deflects electrons from the POI, and the filter attenuates low energy X-rays. Although the acceptance angle of the tracker is already small, a collimator further suppresses interference from photons scattered by surrounding equipments.

\section{Monte Carlo Simulation}
\label{sec:MCS}
Details of the diagnostic process was investigated with Monte Carlo simulations. Interactions of X-rays with a $2$D silicon tracker, the INTPIX5 SOI tracker\cite{miyoshi_monolithic_2013}, which has $\rm 12$ $\rm \mu$m pixel pitch, $\rm 500$ $\rm \mu$m depletion region thickness, and $\rm 1408 \times 896$ pixels was simulated using the Allpix2 simulation package(version 1.6.0)\cite{spannagel_allpix2_2018}. Such simulations mainly comprises two steps. The first step is the particle interaction with the sensor and the generation of charge carriers. This step is implemented with Geant4 toolkit, the G4EmLivermorePhysics in low-energy electromagnetic physics lists was adopted. The second step is the propagation of charge carriers and signal transfer from the sensor to the readout chip. Allpix2 provides different methods to simulate this step, ProjectionPropagation module and SimpleTransfer module were adopted. In the second step, the \textcolor{black}{low limit} threshold was $\rm 200$ e (approximately $\rm 0.73$ keV) and a Gaussian readout noise with a standard deviation of $\rm 50$ e (approximately $\rm 0.18$ keV) was included. This noise level can be achieved on existing silicon-based pixel detectors.

In the investigation, we mainly focused on two issues. The first was the energy resolution of incident X-rays, and the second was the maximal number of tracks that can be identified once a time, which determines the applicable X-ray flux and the accuracy of obtained spectra.

\subsection{Photon energy reconstruction}

\textcolor{black}{Fig.\ref{Fig:FWHMDependence} shows that the angular resolution of electron track algorithm degrades while the electron energy decreases, and therefore the error of inferred incident X-ray energy would be unacceptable for low-energy electrons because they trigger too few pixels.} Therefore, only tracks consist of pixels more than a threshold are considered valid. \textcolor{black}{And the threshold can also restrict the $\rm | \beta |$ angle. The ratio of valid events to detected events is the algorithm efficiency. And the pixels number threshold is chosen as six which is a trade-off between the energy resolution and algorithm efficiency.}

\begin{figure}[!htb]
\centering
	\includegraphics[width=0.8\textwidth]{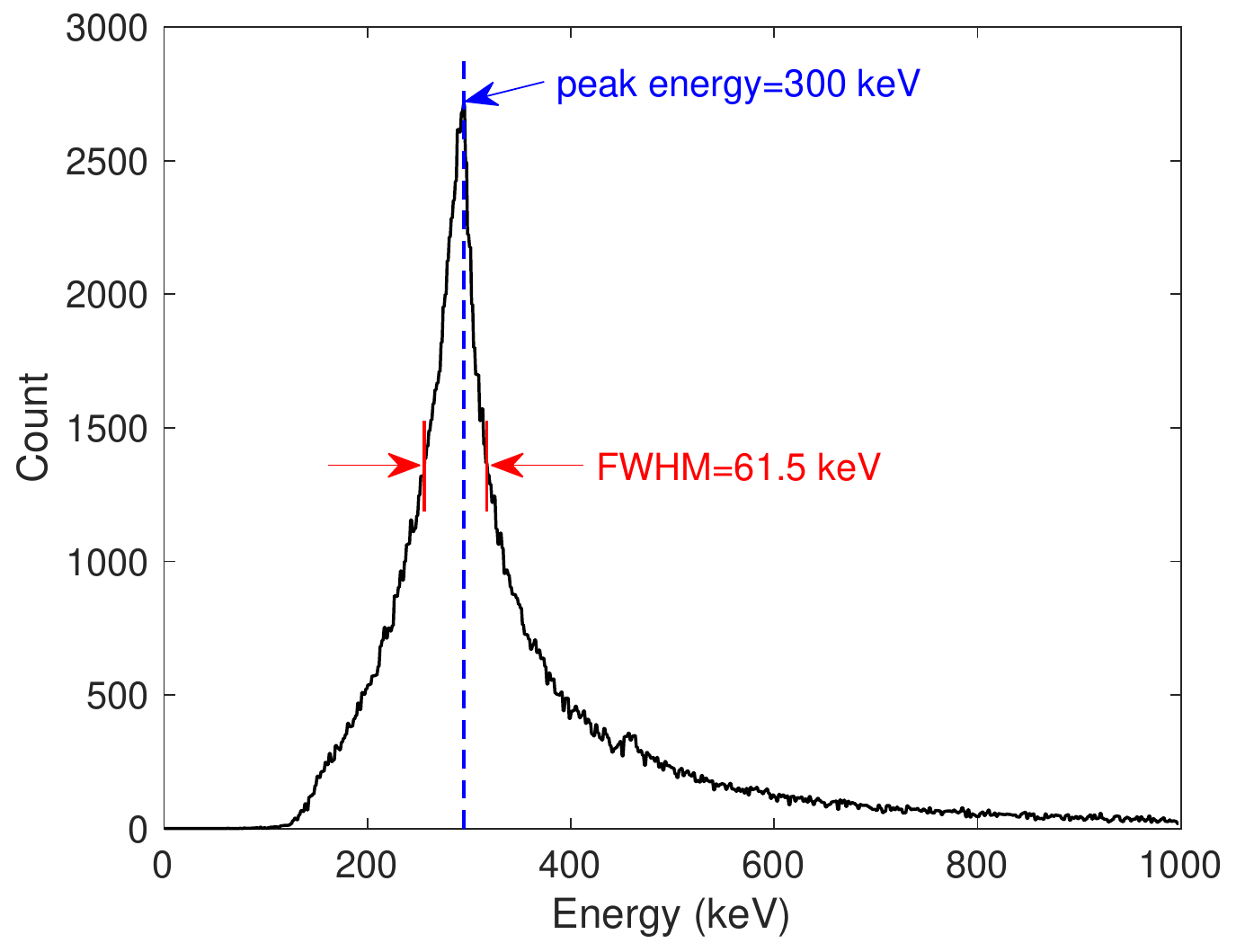}
	\caption{Reconstructed spectrum of a monochromatic source with energy of 300 keV, events with triggered pixel numbers less than six are neglected}
\label{Fig:ReconstructedPhotonSpectrum}
\end{figure}

\begin{figure}[!htb]
\centering
	\includegraphics[width=0.95\textwidth]{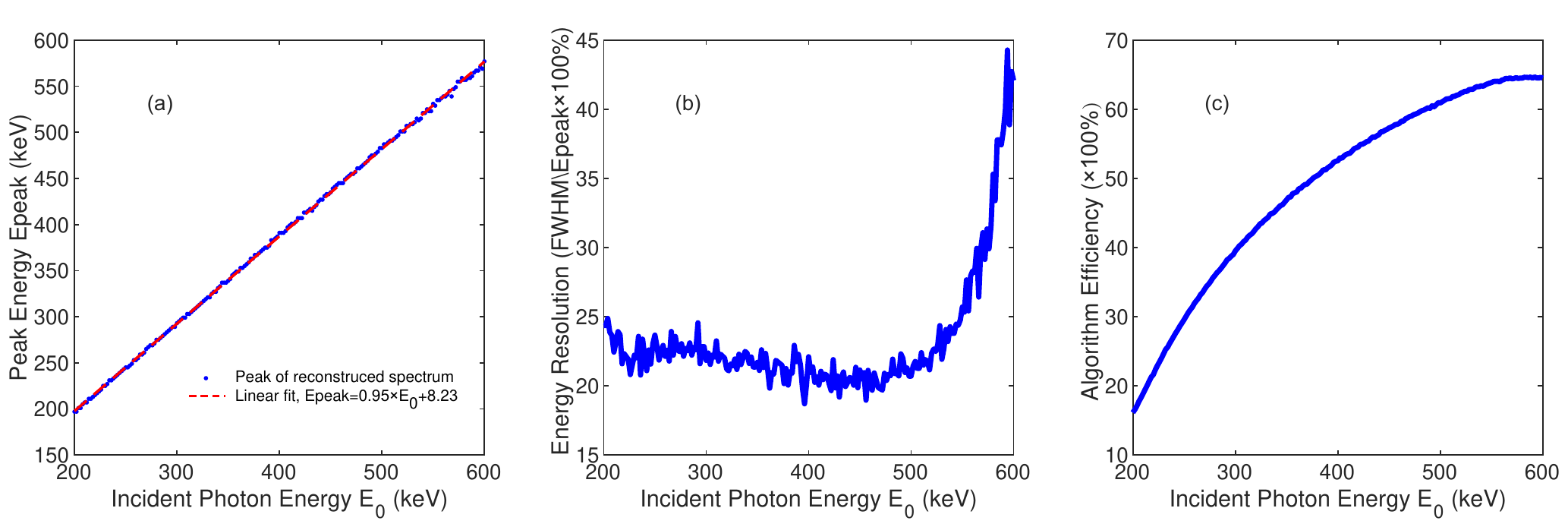}
	\caption{Reconstructed spectrum peak position, energy resolution and algorithm efficiency.}
\label{Fig:EnergyResolutionAndEfficiency}
\end{figure}

Fig.\ref{Fig:ReconstructedPhotonSpectrum} shows the reconstructed spectrum for a monochromatic $\rm 300$ keV X-ray source. Events with triggered pixel number below $\rm 6$ are neglected. The reconstructed spectrum is not Gaussian, and the energy resolution is defined as the FWHM divided by the photon energy. Energy resolution and algorithm efficiency for different incident photon energies are obtained with simulation results and shown in Fig.\ref{Fig:EnergyResolutionAndEfficiency}.

The pixel number threshold is a key parameter that determines the energy resolution and algorithm efficiency. Higher thresholds would filter out more tracks, therefore result in better energy resolution but lower algorithm efficiency. Our choice of this threshold, i. e., $\rm 6$, is a compromise between algorithm efficiency and energy resolution based on detector parameters. With this choice, as shown in Fig.\ref{Fig:EnergyResolutionAndEfficiency}, the energy resolution is better than $\rm 25\%$ and the algorithm efficiency is better than $\rm 10\%$ from $\rm 200$ to $\rm 550$ keV, which can be considered as the measuring range of this diagnostic, and the reconstructed spectrum peak agrees well with the incident photon energy. The reason for the degradation of energy resolution in the high-energy region at the right end of \textcolor{black}{ Fig.\ref{Fig:EnergyResolutionAndEfficiency} (b)} is the escaping of electrons and the approximation of $\rm \beta =0$ no longer stand. In the low-energy region, Compton electrons usually trigger only a few pixels, and the algorithm efficiency is quite low. When the energy of an incident photon is below $\rm 200$ keV, the cross sections of the photoelectric effect and Compton scattering are comparable. Most of the long tracks are induced by more-energetic photoelectrons, resulting in imprecise reconstructions of incident photon energy.

\begin{figure}[!htb]
	\centering
		\includegraphics[width=0.8\textwidth]{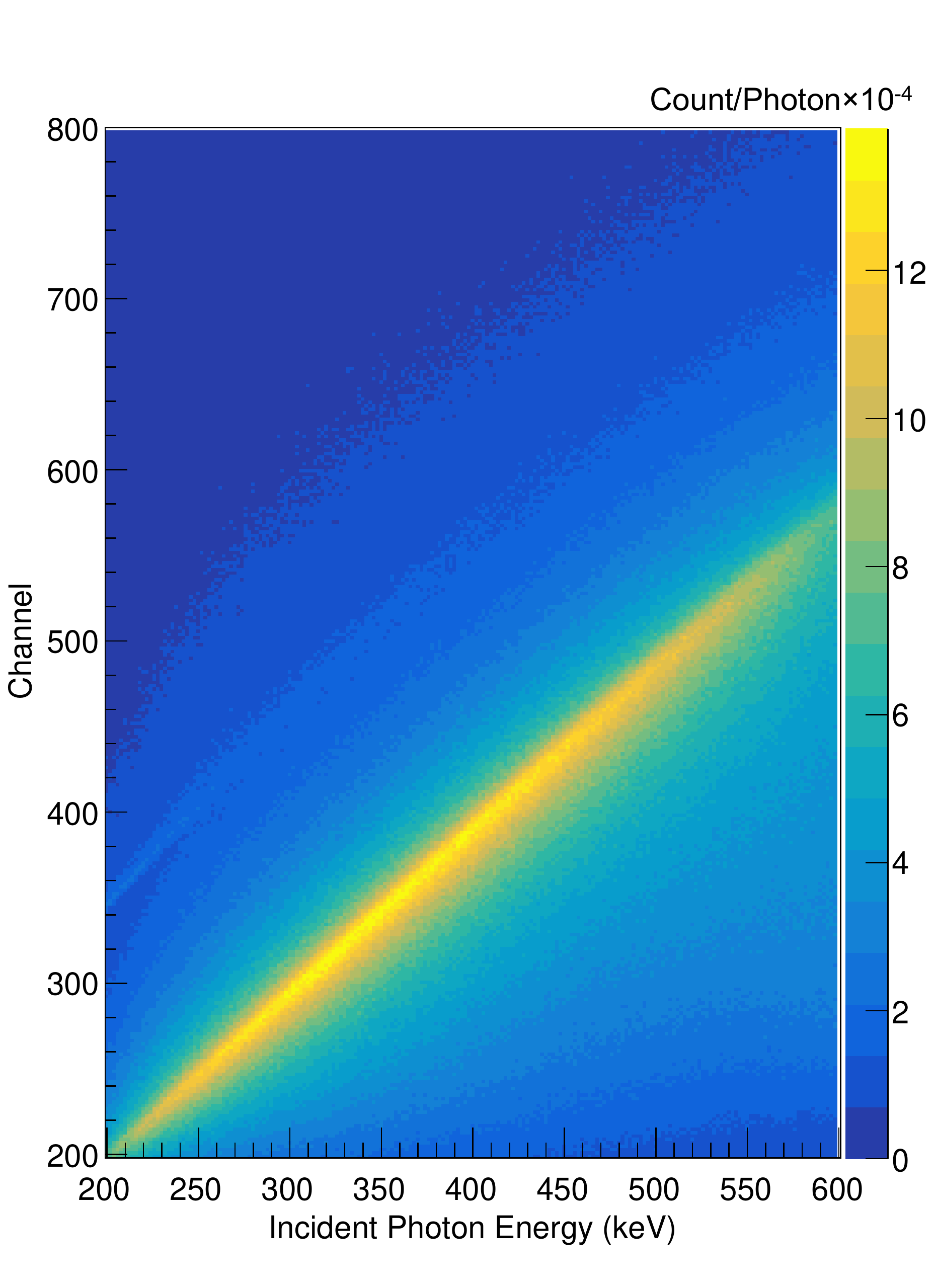}
		\caption{Response matrix of our spectrometer system.}
	\label{Fig:ResponseMatrix}
\end{figure}

The response matrix of this spectrometer system is derived and shown in Fig.\ref{Fig:ResponseMatrix}. The characteristics of this response matrix indicate that the ill conditioning of this system is alleviated, hence the unfolded spectrum would be more accurate.

\subsection{Clustering algorithm and spectrum unfolding}

\begin{figure}[!htb]
	\centering
		\includegraphics[width=0.95\textwidth]{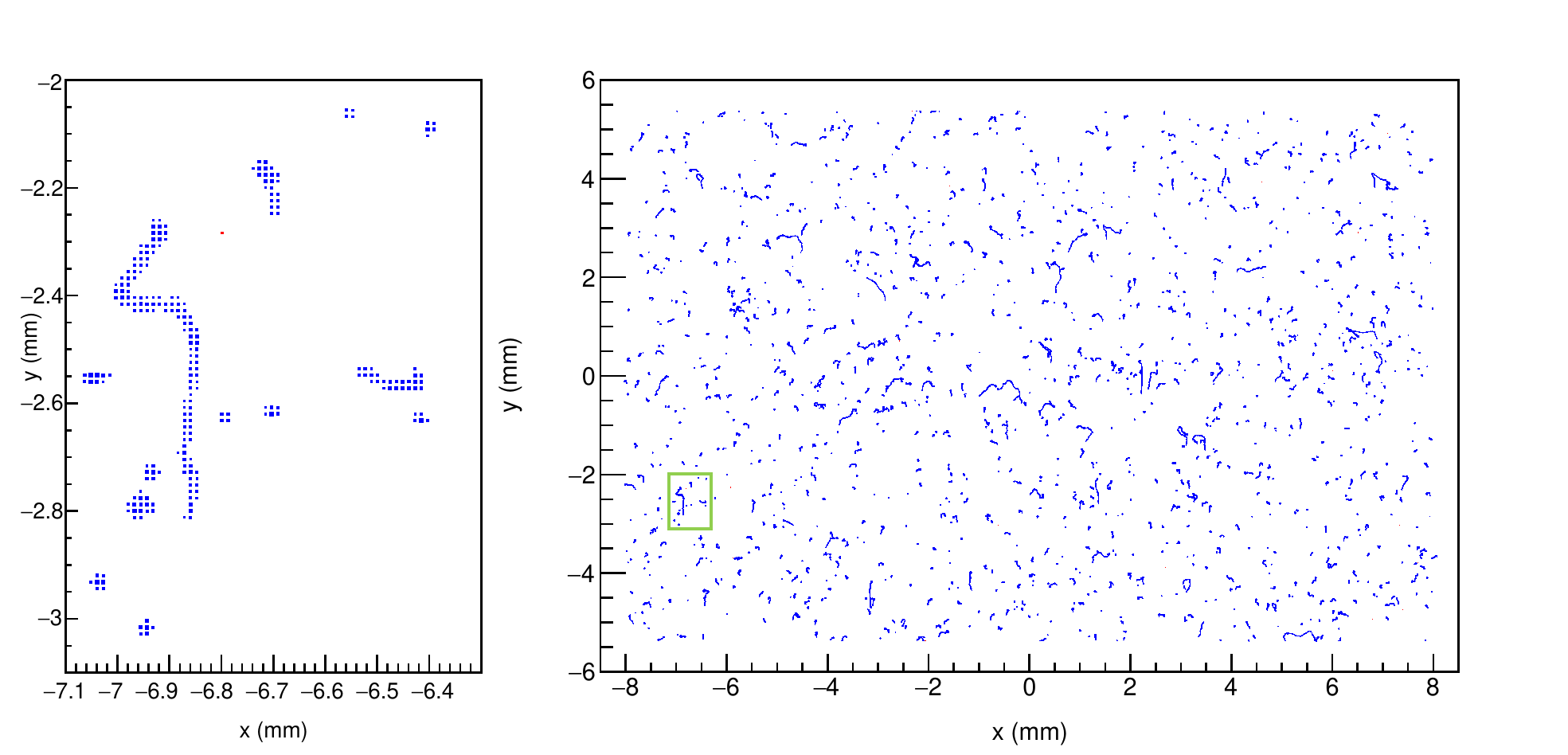}
		\caption{Simulated detector response to an X-ray pulse. Track points are blue and solitary red points are noises identified by the DBSCAN algorithm. On the left shows the details in the green rectangle.}
	\label{Fig:frame_track}
\end{figure}

The pulse durations of laser-plasma X-ray sources are usually very short. Hence, similar to CCDs working at single-photon counting mode \cite{yan_calibration_2013}, a pixel detector generates a frame of gray level image when interacts with a laser-plasma produced X-ray pulse. In this frame of image, multiple tracks recorded as shown in Fig.\ref{Fig:frame_track}, then a clustering algorithm is needed to extract the tracks.

The density-based spatial clustering of applications with noise (DBSCAN) clustering algorithm \cite{schubert_dbscan_2017} was adopted in this study. It can find clusters with any shape and ignore the low-intensity areas as noises. DBSCAN is suitable for processing samples with complex shapes and uncertain cluster numbers, such as electron tracks. 

To simulate the detector response, the total radiation function is modelled by
\begin{equation}
    \rm
    \Phi_{total}(E_\gamma)=\Phi_{Gaus}(E_\gamma)+\Phi_{Brems}(E_\gamma)+\Phi_{BG}(E_\gamma),
    \label{eq:radiation}
\end{equation}
where $\rm \Phi_{Gaus}(E_{\gamma})=\frac{A}{\sqrt{2\pi}\sigma}exp(-\frac{(E_{\gamma}-\mu)^2}{2\sigma^2})$ is a quasi-monoenergetic component, $\rm \Phi_{Brems}=\frac{B}{E_\gamma}(\nu-ln(E_\gamma))$ is the on-axis bremsstrahlung component, and $\rm \Phi_{BG}(E_\gamma)=Cexp(-\eta E_\gamma)$ is the background component\cite{schumaker_measurements_2014}\cite{jeon_broadband_2015} and the unit of $\rm E_\gamma$ is MeV. Parameters of the model radiation spectral function were chosen as $\rm A=2.19\times F$, $\rm B=4.09\times 10^{-3}\times F$, $\rm C=1.28\times F$, $\rm \mu=0.35$, $\rm \sigma=0.02$, $\rm \nu=3.88$ and $\rm \eta=1$. \textcolor{black}{The intensities parameters A, B and C and the shape parameters of bremsstrahlung and background component $\rm \nu$ and $\rm \eta$ were choose according to Ref.\cite{jeon_broadband_2015}. And the mean value and the standard deviation of the quasi-monoenergetic component are set to 350 keV and 20 keV respectively to demonstrate the ability of the spectrometer.} $F$ determines the total intensity, and the total photon number in the range of 200 to 550 keV is $\rm2521 \times F$.

\begin{figure}[!htb]
\centering
	\includegraphics[width=0.8\textwidth]{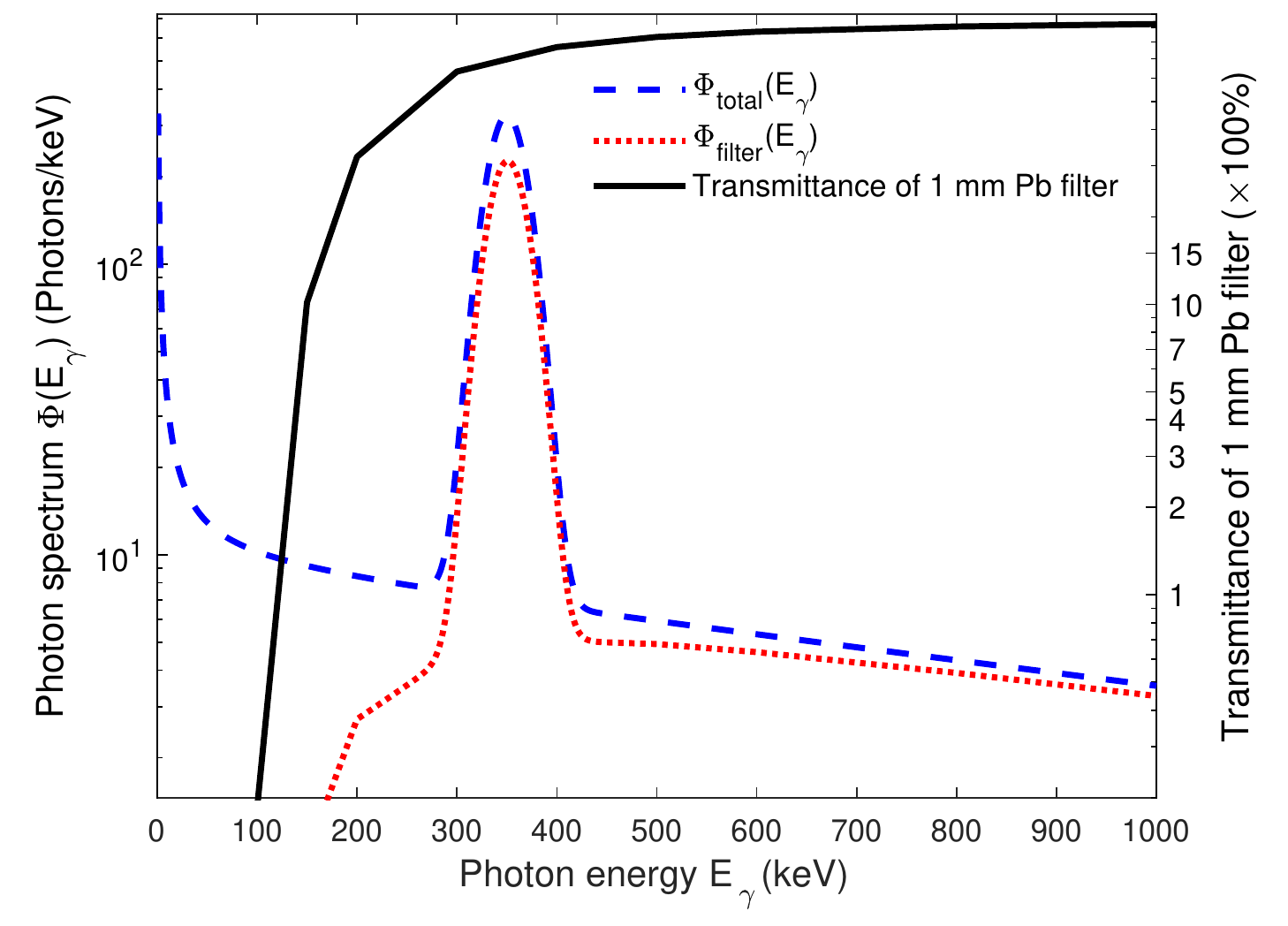}
	\caption{Spectrum modeled by Eq.(\ref{eq:radiation}) (blue dashed line). The transmittance of 1 mm Pb filter (black solid line). And the spectrum incident on the tracker (red dotted line). }
\label{Fig:SpectrumAndEfficiency}
\end{figure}

As a single-shot spectrum diagnostic method, the total number of tracks should be restricted under a certain limit to avoid track overlapping. Therefore, the detection efficiency outside the measuring range, i. e., $\rm 200-550$ keV should be low enough to make sure most detected tracks can be used to reconstruct the X-ray spectrum. A $\rm 1$ mm lead filter is therefore placed between the POI and the detector to filter low-energy photons. Effect of this filter is shown in Fig.\ref{Fig:SpectrumAndEfficiency}, most X-rays below $\rm 200$ keV are absorbed.

\begin{figure}[!htb]
\centering
	\includegraphics[width=0.8\textwidth]{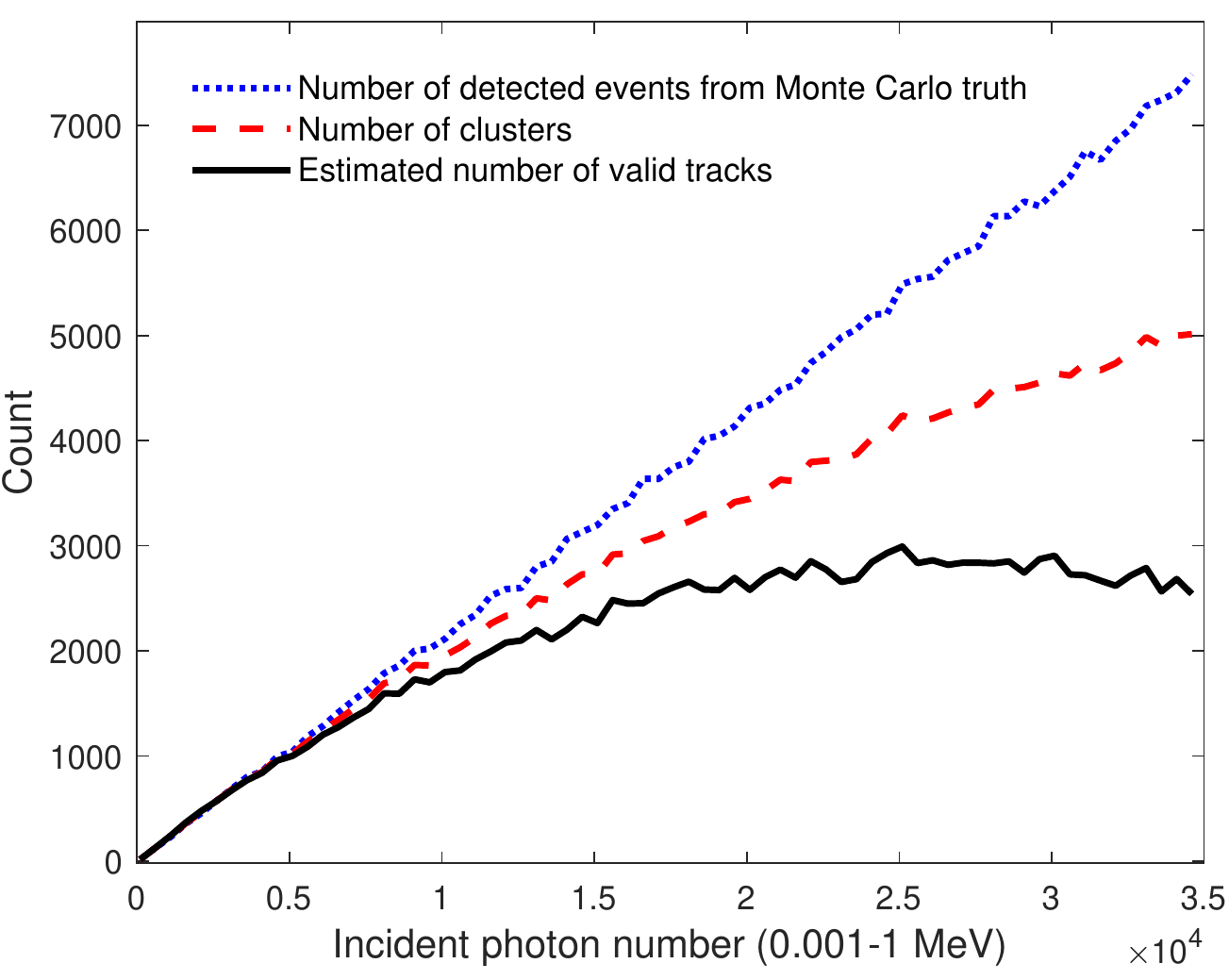}
	\caption{The relation between incident photon number and cluster number.}
\label{Fig:ClusterNumber}
\end{figure}

To make full use of the detector and avoid track overlapping, detector response to X-ray pulses with different fluxes were also investigated. The spectrum of incident X-rays was the same to that in Fig.\ref{Fig:SpectrumAndEfficiency} (with filter). As Fig. \ref{Fig:ClusterNumber} shows, the number of clusters identified by DBSCAN is basically equal to the number of detected events when the incident X-ray number is comparatively low. With the increase of incident photons, the difference between these two numbers, which is caused by track overlapping, grows gradually. The number of valid tracks can be estimated by the number of clusters minus the difference between the number of detected events. \textcolor{black}{Therefore, the optimal number of photons incident on the detector ($\rm 0.001-1$ MeV) is about $\rm 0.5-2.5 \times 10^4$ photons, and the corresponding photon flux is $\rm 0.625-3.125 \times 10^5$ $\rm photons/cm^2$ for the detector with a lateral area of $\rm 0.08\ cm^2$ in simulation.}

The response of a spectrum diagnostic system can be described as \cite{shustov_matrix_2015},
\begin{equation}
    \rm
    Z_i=\sum_{j=1}^{M}\Phi(E_j)K_i(E_j),
    \label{eq:response}
\end{equation}
where $\rm Z_i$ is the spectrum registered by detector channel $i$ ($\rm i=1...N$, where $N$ is the number of channels), $\rm \Phi(E_j)$ is the spectrum corresponding to the incident pulse, and $\rm K_i(E_j)$ is the response matrix of the system, and represents the expectation of detected photon number in channel $\rm i$ when a photon with energy $\rm E_j$ is incident. The response matrix is shown in Fig.\ref{Fig:ResponseMatrix}. The spectrum unfolding is based on the expectation maximisation method\cite{zhang_x-ray_2007}, it is applied to solve the inverse problem in Eq.(\ref{eq:response}) with known $\rm K_i(E_j)$ and $\rm Z_i$ to obtain $\rm \Phi(E_j)$.

\begin{figure}[!htb]
\centering
	\includegraphics[width=0.8\textwidth]{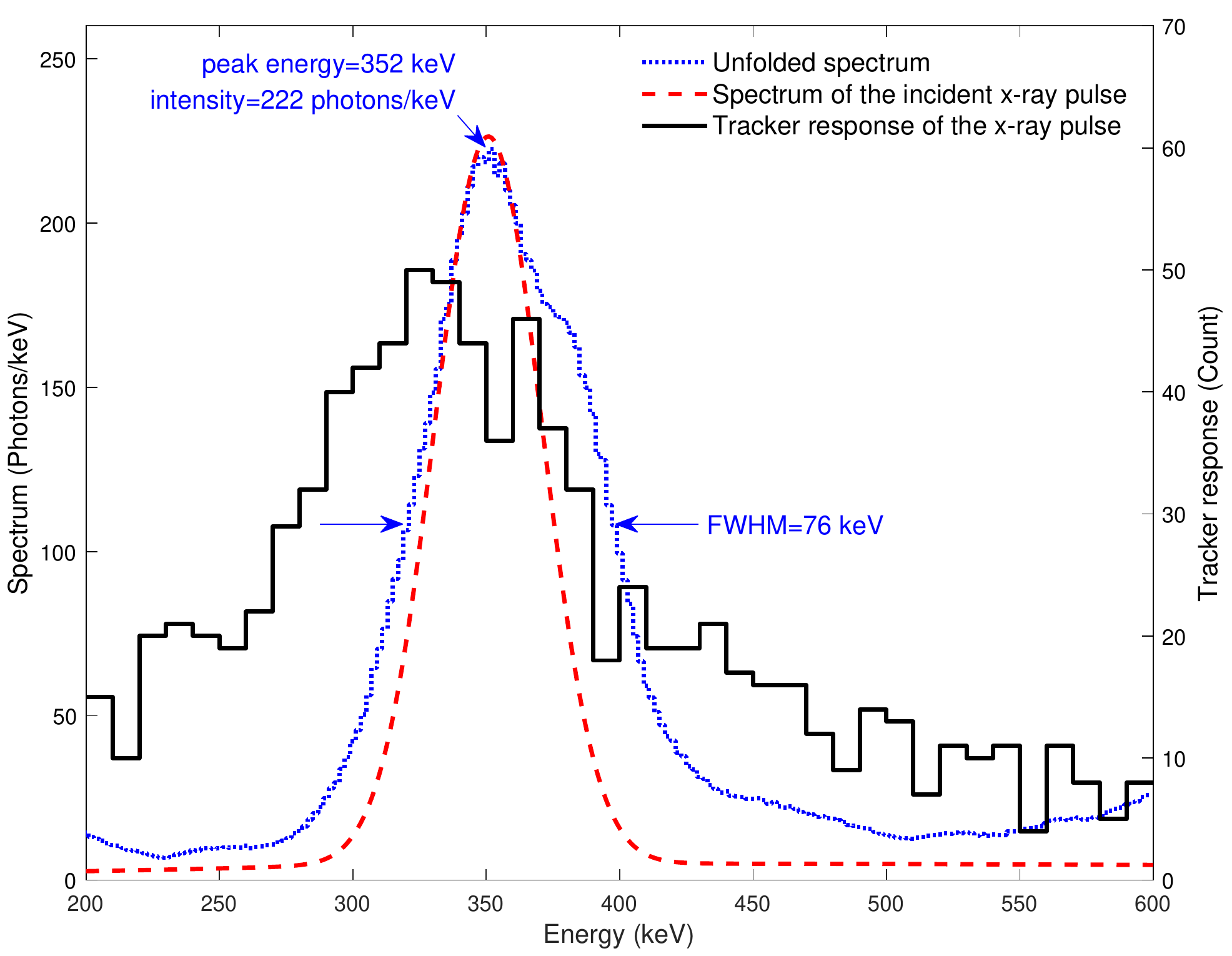}
	\caption{Tracker response and unfolded spectrum}
\label{Fig:UnfoldSpectrum}
\end{figure}

An example to unfold the spectrum was carried out. In this example, the spectrum of incident X-ray pulse was the same to that in Fig.\ref{Fig:SpectrumAndEfficiency}, \textcolor{black}{the quasi-monoenergetic component had a peak energy of 350 keV, FHWM of 47.1 keV and peak intensity of 226 photons/keV.} Simulation gave the response of the detector, then we unfolded the X-ray spectrum with the method above. The incident spectrum, unfolded spectrum, and the response of tracker are shown in Fig.\ref{Fig:UnfoldSpectrum}. \textcolor{black}{We find the peak energy, FWHM and peak intensity of the unfolded spectrum are 352 keV, 222 photons/keV and 76 keV and the relative deviations are $\rm 0.5\%$, $\rm -1.7\%$ and $\rm 61.7\%$ respectively. The peak energy and intensity of unfolded spectrum agree well with those of the incident spectrum. The measurement accuracy of FWHM is limited by the energy resolution and the statistical uncertainties, which can be further improved by increasing the number of trackers employed in one spectrum diagnostic system.}

\section{Proof-of-Principle Experiment}
\label{sec::results}
A proof-of-principle experiment was conducted using a hybrid pixel detector Timepix3. Timepix3 \cite{poikela_timepix3_2014} is a hybrid readout chip that contains $\rm 256 \times256$ pixel channels ($\rm 55\ \mu$m $\rm \times 55\ \mu$m each). Timepix3 can measure the charge via Time-over-Threshold and the depth of interaction position via time-of-arrival in each pixel simultaneously, and can provide the possibility of the three dimensional (3D) particle trajectory reconstruction or 3D interaction position reconstruction. A depth resolution of $\rm \sim 50\ \mu$m can be achieved for a $\rm 500\ \mu$m silicon sensor at $\rm 130$ V bias voltage after time-walk correction\cite{bergmann_3d_2017}. \textcolor{black}{The Timepix3 detector we used is MiniPIX TPX3 manufactured by ADVACAM and equipped with a $\rm 500$ $\rm \mu$m silicon sensor. The maximal pixel rate of MiniPIX TPX3 is 2.35 M pixels per second.}

\begin{figure}[!htb]
	\centering
		\includegraphics[width=0.8\textwidth]{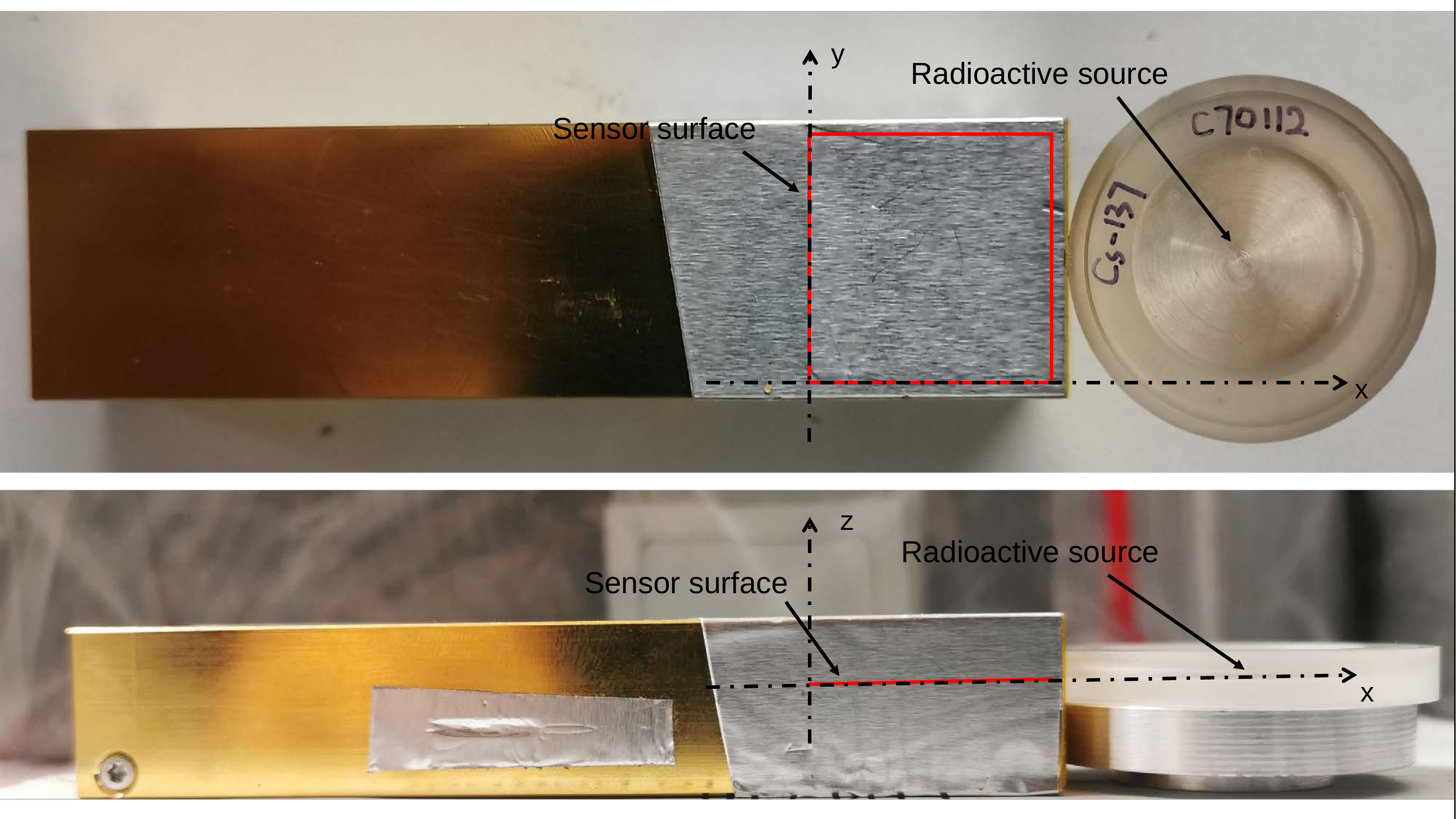}
		\caption{Setup of proof-of-principle experiment. Upper panel is a bird view and lower panel is a side view.}
	\label{Fig:ExpSetup}
\end{figure}

\begin{figure}[!htb]
\centering
	\includegraphics[width=0.8\textwidth]{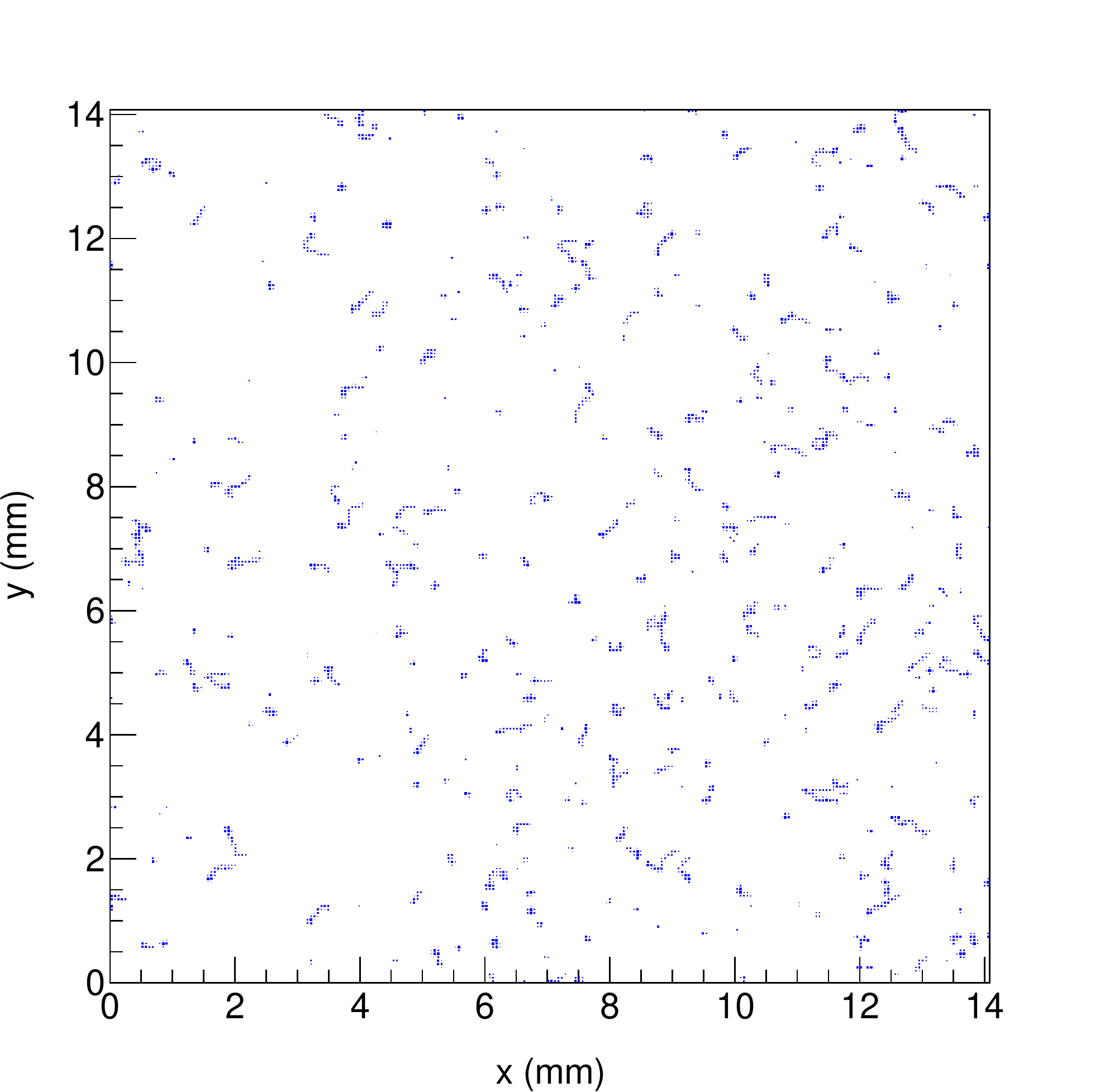}
	\caption{Electron tracks obtained in the experiment.}
\label{Fig:Timepix3Response}
\end{figure}

The experimental setup and the definition of coordinates are shown in Fig.\ref{Fig:ExpSetup}. The Timepix3 was operated at a bias voltage of $\rm 130$ V. The energy resolution of the Timepix3 detector at 59.5 keV is 2.0 keV (standard deviation), and the calibration gives the depth resolution to be $\rm 60 \mu$m. The sensor of Timepix3 is a square located at $\rm 0<x<14$ mm, $\rm 0<y<14$ mm, and $\rm 0<z<0.5$ mm. It is irradiated by a $\rm ^{137}$Cs radioactive source with $\rm 661.5$ keV gamma-rays placed at $\rm (27, 7, 0)$ mm. The goal of the proof-of-principle experiment is to verify the feasibility of reconstruction X-ray spectrum from Compton electron track, therefore, the activity of the $\rm ^{137}$Cs source used is only $\rm 1000$ Bq and the count rate of the Timepix3 detector is approximately 5 pixel per second. The detector was placed in a lead shield to depress the environmental radiological background. The detector response of $\rm 300$ events was accumulated and plotted in $\rm 2$D and shown in Fig.\ref{Fig:Timepix3Response}. As discussed in Ref.\cite{plimley_angular_2016}, detectors with a pixel pitch larger than $\rm 40\ \mu$m have poor angular resolution for electrons of several hundred keV. Hence a track pixel number threshold of $\rm 15$ was adopted to select longer tracks and realize better angle resolution at the expense of efficiency.

\begin{figure}[!htb]
	\centering
		\includegraphics[width=0.8\textwidth]{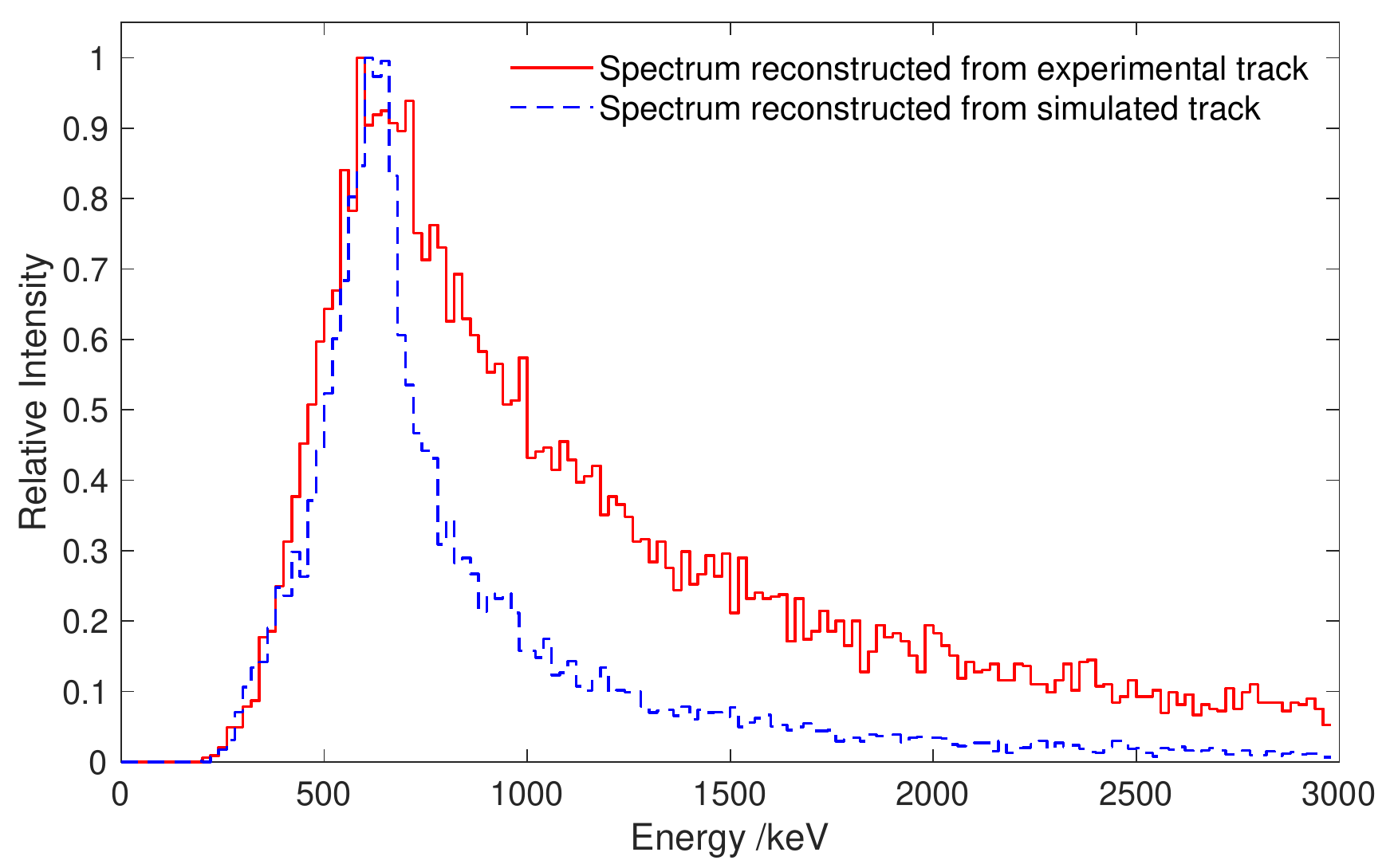}
		\caption{Spectra reconstructed in the experiment and simulation.}
	\label{Fig:ExperimentSpectrum}
\end{figure}

A $\rm 3$D electron track algorithm was used in this experiment to determine the interaction point and scattering direction of Compton electrons. Then the incident photon spectrum was reconstructed with a known radioactive source location. A $\rm 3$D Monte Carlo simulation consistent with the experimental setup was also conducted. The photon spectra reconstructed from the simulated and experimental tracks are shown in Fig.\ref{Fig:ExperimentSpectrum}. Reconstructed spectral peak is approximately at $\rm 661.5$ keV, which is in good agreement with the simulation results and the radioactive source. \textcolor{black}{Broadening of the experimental photon spectrum relative to the simulated photon spectrum is mainly caused by two reasons. First, the effective area of the copper housing of Timepix3 is much larger than that of the sensor which means the photons scattering in the copper housing can have a distinct contribution to the experimental spectrum. But the copper housing is ignored in the simulation. Second, the $\rm ^{137}$Cs is not a point-like source, and the spatial distribution of $\rm ^{137}$Cs is uncertain. The spread of $\rm ^{137}$Cs source will lead to a greater error of the Compton scattering angle $\rm \phi$ in the experiment, but a perfect point $\rm ^{137}$Cs source is used in the simulation}.

\section{Conclusion and Outlook}
A single-shot spectrum diagnostic method was proposed for an ultrashort X-ray pulse in the few hundred keV to sub-MeV band. In this method, semiconductor pixel trackers record projected tracks of Compton scattered electrons. With clustering algorithm and electron track algorithm, energies and initial directions of Compton electrons can be recovered from obtained Compton electron tracks to reconstruct the X-ray spectrum. Since this is a single-photon counting diagnostic method without converter, the photon spectrum can be measured with high efficiency and reasonable energy resolution, which cannot be achieved by existing diagnostic technologies in this energy band. \textcolor{black}{As an on-line diagnostic method, it is suitable for high repetition rate laser-plasma-based X-ray sources.}

Simulations show that such diagnostic based on a $\rm 2$D silicon tracker with a pixel pitch of $\rm 12\ \mu$m and a thickness of $\rm 500\ \mu$m can achieve a $\rm 25 \%$ energy resolution from $\rm 200$ to $\rm 550$ keV. \textcolor{black}{The upper limit of valid event number detected for a typical detector with $\rm \sim 10^6$ pixels was approximately a few thousand, and the optimal photon flux is $\rm 0.625-3.125 \times 10^5$ $\rm photons/cm^2$ for a typical detector with a lateral area of $\rm 0.08\ cm^2$.} A proof-of-principle experiment based on a Timepix3 hybrid detector was carried out, and the spectrum of a $\rm 661.5$ keV radioactive source was reconstructed with approximately $\rm 60 \%$ energy resolution.

With the development of high-pixel-density $\rm 3$D trackers and the improvement of sensor thickness, the energy resolution and applicable energy range can be further improved. The total number of valid Compton tracks is limited by the pixel number of the detector, hence the statistical stability of reconstructed X-ray spectrum can be improved by increasing the number of detectors.

\section{Acknowledgements}
This work was supported by the National Natural Science Foundation of China (Grant No. 11975214 and 12004353), Science Challenge Project (Grant No. TZ2018005) and National Key R\&D Program of China (Grant No. 2016YFA0401100).

\bibliography{bibfile}

\end{document}